# Introducing E-Maintenance 2.0

**Abdessamad Mouzoune**
QSM Laboratory - Ecole Mohammadia d'ingénieurs
Mohammed V University – Agdal Rabat, Morocco

**Saoudi Taibi**
QSM Laboratory - Ecole Mohammadia d'ingénieurs
Mohammed V University – Agdal Rabat, Morocco

**ABSTRACT**
While research literature is still debating e-maintenance definition, a new reality is emerging in business world confirming enterprise 2.0 model. Executives are more and more forced to stop running against current trend towards social media and instead envisage harnessing its power within the enterprise. Maintenance can't be an exception for long and has to take advantage of new opportunities created by social technological innovations. In this paper a combination of pure « e » perspective and « 2.0 » perspective is proposed to avoid a lock-in and allow continuous evolution of e-maintenance within the new context of business: A combination of data centric models and people oriented applications to form a collaborative environment in order to conceive and achieve global goals of maintenance. New challenges are also to be expected as to the efficient integration of enterprise 2.0 tools within current e-maintenance platforms and further research work is still to be done in this area.

## 1.    INTRODUCTION

The evolution of maintenance is naturally seen through the scope of the evolution of industrialization itself, its mechanization and its automation. Moubray traced the resulting evolution through three generations [1]. First Generation: Within the period up to World War II industry was not very highly mechanized and most equipment was simple and over-designed with no significant need to worry about the prevention of equipment failure. Systematic maintenance was mainly about simple cleaning and lubrication routines with lower need for skills. Second Generation: Increased mechanization and more complex equipment have made from downtime a real concern bringing more focus to means and concepts that would prevent equipment failures. Preventive maintenance in the sixties was principally led as periodic general revisions of   equipments. In addition to control systems, this period also knew a significant trend toward maintenance planning to control maintenance costs while trying to increase and take full advantage of the life of the assets. The Third Generation: The new expectations that have marked this period starting from the middle of the 70's due to the acceleration of the change in industry were mainly: Condition monitoring, Design for reliability and maintainability, Hazard studies, Small fast computers, Failure Mode and Effect Analysis, Expert systems, Multi-tasking and teamwork. Maintenance techniques that were developed in this period such as FMEA have proven their suitability in many critical fields including mobile health monitoring systems [2] where proper functioning is of critical importance for the safety of patients.

In manufacturing, impacts of downtime are strengthened by the world wide adoption of just-in-time systems. In this context, automation has the potential to connect engineering design, manufacturing and enterprise systems, enabling a customer-driven, responsive production environment. With emerging applications of Internet, communication technologies and the impact of e-intelligent paradigm [3], companies change their manufacturing operations from local factory integration and automation to global enterprise automation with the ability to exchange information and synchronize with different e-business systems [4].

In these circumstances, the concept of e-maintenance emerged as a result of the integration of ICT technologies in maintenance policies to deal with new expectations of innovate solutions for e-manufacturing and e-business [5].

In section 2, we describe and motivate the problem we are going to consider under the new reality set up by business 2.0 model. In section 3 and 4, we will study respective characteristics of "E" and "2.0" perspectives and propose their combination in section 5 to end with conclusion.

## 2. SETTING THE PROBLEM

Interested in general approaches, we gathered 107 publications for the period from 2000 to the end of 2013 using internet research (Google Scholar, IEEE Xplore ...) against the word « e-maintenance » in title or keywords. A summary study showed us that « E » Perspective is unanimously accepted: Within such a perspective, e-maintenance is explicitly or implicitly included in a natural scope of E-enterprise that is an instantiation of the e-business concept at the level of an enterprise.

However, Enterprise 2.0 is another emergent scope that is radically changing the world of doing business. While Section 4 will cover this "2.0" perspective, let us mention for now that maintenance managers are already influenced by diverse 2.0 technologies and use them in a large amount of their communications with all members of their staff and more often beyond formal and secure IT systems. Instant messaging and wikis are examples of such tools that can enhance organizational communication if well deployed within an enterprise.

In this paper we are interested in the question of how the evolution of e-maintenance concept can be seen within this new reality as imposed by the « 2.0 » Perspective. We are especially focusing on the main characteristics that distinguish the two perspectives as regarded to their respective data-or-people founding models. In addition to that central founding characteristic, we shall consequently consider collaboration and intelligence in this study.

From all publications we gathered in October 2013, very few were really general articles from which we selected article [5] for its exhaustiveness. While reviewing works in the field of e-maintenance for the first half of the period we are considering, the selected paper is also the most cited general article. Hence, we consider it has most influence on publications of the second half of the period. The authors are also largely known for their contributions within the e-maintenance community.

Although the definition of e-maintenance is still debated by researchers as in [6], we retain the following definition that is proposed in the selected article as it is the first tentative to federate a large number of known definitions: "*Maintenance support which includes the resources, services and management necessary to enable proactive decision process execution. This support includes e-technologies (i.e. ICT, Web-based, tether-free, wireless, infotronics technologies) but also, e-maintenance activities (operations or processes) such as e-monitoring, e-diagnosis, e-prognosis, etc*" [5].

## 3. THE CURRENT « E » PERSPECTIVE

The common term in maintenance-related literature "e-maintenance" is introduced in 2000. In [5], the authors discussed the emergence of e-maintenance concept as a maintenance strategy, as maintenance plan, as maintenance type and as maintenance support while considering it as a key element of the e-enterprise.

Citing [7], the e-enterprise is seen as a combination of ''point-and-click'' net business models and traditional ''brick-and-mortar'' assets leading to next-generation organizations. The authors cite four characteristics that are (1) real-time reaction to customer's demand; (2) an iterative learning approach; (3) holistic methodologies to define each constituent of the enterprise architecture; and (4) alignment of technological choice with the business model.

For our part, we note the following characteristics regarding the « E » perspective:

### 3.1 Data centric models

By considering e-maintenance as part of the e-enterprise, the key words are then integration, openness and interoperability [8]. Data are at the centre of that integration endeavor including standards development such as MIMOSA [9] and platforms development in e-maintenance such as PROTEUS [8], DYNAMITE [10] and TELMA [11].

Those data centric models follow more or less OSA-CBM architecture [12] that the authors of the selected article described in terms of several successive layers: Data Acquisition, Data Manipulation, Condition Monitor, Health Assessment, Prognostics, Automatic Decision Reasoning and Human-Computer Interface.

### 3.2 Collaboration is about sharing information

For the most, definitions that are proposed for e-maintenance consider explicitly or implicitly that collaboration is about sharing information. As an example we cite H. K. Shivanand & al. [13]: *"It is a network that integrates and synchronizes the various maintenance and reliability applications to gather and deliver asset information where it is needed, when it is needed."*

At the best, collaboration is considered as a synchronized and coordinated form of cooperation. When defining collaborative maintenance for instance, the authors of the selected paper give examples such as on-line condition-based monitoring and real-time process monitoring.

### 3.3 Intelligence is about automation

Some definitions clearly link intelligence to automation such as Zhang et al. [14] considering that e-maintenance combines Web service and agent technologies to endow the systems with intelligent and cooperative features within an automated industrial system.

Crespo Marquez and Gupta [15] consider e-maintenance as an environment of distributed artificial intelligence. Each time the authors of the selected article qualify as intelligent a device or a task such as intelligent predictive maintenance, that qualification mainly means its automation. We note that we studied intelligence in current e-maintenance conception and its tendency to automation more extensively in a previous article [16].

### 4. THE INELUCTABLE « 2.0 » PERSPECTIVE

In 2006, Andrew McAfee [17] coined the term "Enterprise 2.0" as *"the use of emergent social software platforms within companies, or between companies and their partners or customers."*

The potential significance of Enterprise 2.0 and other related concepts and products (Social Business and Enterprise Social Software) over the next years in terms of global market is forecast to grow from $US721.3 million in 2012 to $US6.18 billion in 2018 according to Markets-and-Markets 2013 as cited in [18].

However, e-maintenance community does not seem to address significant interest to the emergence of the concept of Enterprise 2.0. David Andersson

[19] mentions some other reasons that enterprise 2.0 is of great importance in relation to enterprise system:

- Communications are already held in companies by means of social media and outside the boundaries of enterprise systems. Such a situation where no record is kept within the system also represents a great issue in terms of security.
- Social media tools offer very helpful technologies to capitalize knowledge within the enterprise concerning its equipments and its processes. Experts' knowledge is then preserved in order to be used by other people in the company even when initial experts leave it.
- Dynamic formats such as wikis to document current processes as well as their changes over time are a way to improve complex front office processes (e.g. Engineering, Project management and others).

Peter Drucker predicted that competitive and participative environment was leading working groups' members to become what he called "knowledge workers." He goes further, arguing that each knowledge worker whose contribution affects the performance of the organization is an "executive" [20]. Hence, considering that most staff in maintenance if not all are "knowledge workers", we list characteristics of the « 2.0 » perspective as follows:

### 4.1 People oriented applications

Andrew McAfee created the acronym "SLATES" about the use of social software within the context of business. Each of the following six components of the SLATES acronym standing for main people oriented applications provides an essential component of Enterprise 2.0 as cited in [21]:

- Search: Knowledge workers in maintenance would be able to find what they are looking for inside the company or outside via internet by using personalized and more efficient keywords;
- Links: Links are one of the key indicators that search engines use to assess the importance of content in order to deliver accurate and relevant results. They also provide guidance to knowledge workers about what is valuable;
- Authoring: The intranet would be no more created by a restricted number of people to become a dynamic support of collective knowledge if employees are given the tools to author information;
- Tags : By allowing knowledge workers to attach tags to the information they create and find valuable, taxonomies emerge based on actual practice which is to help information architects to organize information by meaning;

- Extensions: Tags, authoring and links would allow knowledge engineers to identify patterns and use these as extensions to information and relationships.
- Signals: Technologies such as really simple syndication (RSS) allow employees to efficient use information in a controlled way.

### 4.2    Collaboration is about expertise sharing:

The importance of collaboration is mentioned in media and literature regarding Web 2.0 such as Hinchcliffe in [22] arguing that enterprise 2.0 and Web 2.0 is about new forms of collaboration and communities "not communities' new plumbing."

Since within e-maintenance literature, collaboration is often used as a form of cooperation, it is of great interest to emphasize their differences in the context of enterprise 2.0 as reminded in [21]:

Cooperation is based on the division of labor, each person responsible for portion of work while tasks are split into independent subtasks and coordination is only required when assembling partial results. Cooperation is informal and for short term with no jointly defined goals. Individuals retain authority and information is shared just as needed.

By contrast, collaboration necessitates long term mutual engagement of participants in a coordinated effort to solve a problem and cognitive processes are divided into intertwined layers. Commitment and goals are shared and so are risks and rewards while collaborative structure determines authority.

In the context of maintenance, collaboration technologies enable members to communicate and collaborate as they deal with the opportunities and challenges of asset maintenance tasks as mentioned in [23].

Expertise location capability is another concept related to this 2.0 perspective enabling companies to solve business problems that involve highly skilled people or when those problems hardly lend themselves to explicit communication [24].

In this orientation, expertise sharing is considered a new metaphor in knowledge management evolution focusing on the inherently collaborative and social nature of the problem [25].

### 4.3    Intelligence is a collective emergent property:

Harnessing collective intelligence is one of the eight principles of Web 2.0 that are described by O'Reilly in [26] where the author mentions its three aspects: (1) Peer Production without traditional hierarchy,  (2) The Wisdom of crowds where large groups of people outperform elite. (3) Network effects from user contributions while sharing added value with others.

## 5. COMBINING PERSPECTIVES FOR E-MAINTENANCE 2.0

In this section, we propose to combine the two perspectives for following reasons:

- To recognize that E-business is changing to new business models within what is called e-business 2.0 [27];
- To consider a combination of pure « e » perspective and the « 2.0 » perspective as an evolution to take advantage of new opportunities created by technological innovations while expecting new challenges such as security;
- To add capitalization of informal and/or implicit knowledge to capitalization of formal and/or explicit knowledge;
- To combine pure e-maintenance capabilities with social technologies and people oriented collaborative applications and platforms within each of maintenance services and tasks such as Condition Monitoring, Diagnostics and Prognostics.

After having extracted main characteristics that differentiate both perspectives in sections 3 and 4, we can combine those extracted characteristics to construct a definition of e-maintenance 2.0 as follows while Figure 1 illustrates this construction:

"*A combination of data centric models and people oriented applications to cooperatively and collaboratively share information and expertise in order to conceive and achieve global goals of maintenance through automation and human intervention.*"

To avoid auto definition, the terms "e" and "2.0" are intentionally omitted in the proposed definition where the term "maintenance" keeps its standard definition. According to the European Standard EN 13306 -2001, the goals of all technical and managerial actions of maintenance are retaining an item in, or restoring it to, a state in which it can perform the required function. Such goals are to be pursued during the whole life cycle of each item.

Global goals of maintenance extend maintenance goals at the scale of the enterprise while insuring strategic alignment with its other constituents and departments.

Automation of maintenance activities reflects the current e-maintenance orientation based on data centric models with or without human intervention.

Cooperation and collaboration are both evoked in order to keep their distinction very explicit while collective emergent property of intelligence is implicit and required not only to achieve goals but to conceive them as well.

Information (the know what) and expertise (the knowhow and know why) form the specter of knowledge and sharing them implies the necessity of the presence of more than one actor be it human or machine. The more actors join, the wider is the consequent network. Data sources are at the core of the knowledge process and get richer as more people use them.

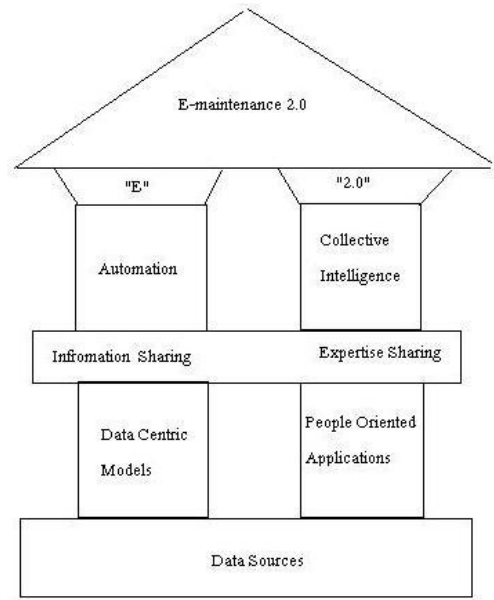

**Figure 1. Illustration of e-maintenance 2.0 construct**

At the end of this section, we note that the combination of the two former perspectives into new one creates new challenges such as:

- Security: To avoid compromising critical information by social media, a high level of importance is to reserve to the ability to ensure that critical information and content of internal conversations is not accessed by unauthorized people.
- Misdirection: Building social media functionalities like instant messaging or wikis within an enterprise platform should ensure that employees remain more productive and don't leave their working context by using web 2.0 tools.
- Integration: New challenges are also to be expected as to efficient technical integration of enterprise 2.0 tools and further research work is still to be done in this area [18].

In our current research, we are considering to deal with this issue within a project we called "Social CMMS": It is an "e-CMMS 2.0" where a known CMMS that is linked to some condition monitoring e-technologies and associated with a collaborative platform as an internal social network offering all SLATES components: The purpose is to explore at which level informal knowledge can be integrated to enhance different services of e-maintenance while following a framework we proposed in [16].

## 6. CONCLUSIONS

This paper presents an overview of the evolution of the e-maintenance concept within current research literature. It retraces the historical path that the concept walked depending on the evolution of industrialization, its mechanization and automation. This kind of path dependency evolution is leading the concept to a lock-in forced by the e-enterprise perspective. A selective review of literature allowed us from one side to confirm the lock-in coming to prominence and, from the other side, to extract main characteristics of the "e" perspective: (1) data centric models (2) collaboration is about sharing information and (3) Intelligence is about automation. To allow the concept of e-maintenance to face the new reality of enterprise 2.0 as it is emerging in business world, we first exposed main characteristics of the new "2.0" perspective : (1)people oriented applications (2) Collaboration is about sharing expertise and (3) intelligence is a collective emergent propriety. After explode extracting main characteristics of both perspectives, a reconstruction of the new concept through a combination of respective characteristics within e-maintenance 2.0 is proposed. We considered the combination of pure « e » perspective and the « 2.0 » perspective as a necessary evolution to take advantage of new opportunities created by social technological innovations, e.g. adding capitalization of informal and/or implicit knowledge to capitalization of formal and/or explicit knowledge- while expecting new challenges such as security. New challenges are also to be expected as to the efficient integration of enterprise 2.0 tools within current e-maintenance platforms and further research work is still to be done in this area.

This paper may be cited as:

Mouzoune, A. and Taibi, S., 2014. Introducing E-Maintenance 2.0. *International Journal of Computer Science and Business Informatics, Vol. 9, No. 1, pp. 80-90*.